\documentstyle[12pt]{article}

\newcommand{\reseteqnum}{\setcounter{equation}{0}}

\newcommand{\be}{\begin{equation}}
\newcommand{\ee}{\end{equation}}
\newcommand{\bea}{\begin{eqnarray}}
\newcommand{\eea}{\end{eqnarray}}
\newcommand{\nn}{\nonumber\\}

\begin{document}

\begin{titlepage}
% Preprint Numbers
\begin{flushright}
hep-th/0009043 \\
KEK-TH-710
\end{flushright}
\bigskip
\bigskip

% Title
\begin{center} \Large\bf 
  Nonperturbative Aspect
  In ${\cal N}=2$ Supersymmetric \\ 
  Noncommutative Yang-Mills Theory
\end{center}
\bigskip

% Author
\begin{center} \large
   Yuhsuke Yoshida
\end{center}

% Address
\begin{center} \normalsize \it
   KEK Theory Group,
   Tsukuba, Ibaragi 305-0801,
   JAPAN \\
{\rm e-mail : yoshiday@post.kek.jp}
\end{center}
\bigskip

% Abstract
\begin{center} 
\Large \bf Abstract
\end{center}
\begin{quote}
We investigate asymptotic behaviors of the strong coupling limit
in the ${\cal N}=2$ supersymmetric non-commutative Yang-Mills theory.
The strong coupling behavior is quite different from the commutative
one since the non-commutative dual $U(1)$ theory
is asymptotic free, although the monodoromy is the same as
that of the ordinary theory.
Singularities are produced by infinitely heavy monopoles and dyons.
Nonperturbative corrections may be determined by holomorphy.
\end{quote}

\end{titlepage}

%%%%%%%%%%%%%%%%%%%%%%%%%%%%%%%%%%%%%%%%%%%%%%%%%%%%%%%%%%%%%%%%%%%%%%%%%%%%%

\section{S-duality of Non-commutative $U(1)$ gauge theory}\label{SD}
\reseteqnum

In this note we consider the non-perturbative aspect of non-commutative
Yang-Mills theory making good use of supersymmetry and duality.
We also use perturbative analyses done in Refs.\cite{MRS,Haya,RS}. 
Strong coupling region can be analyzed by using duality.
Holomorphy puts severe constraints on the funcitonal form of the
prepotential.
We show a strong coupling behavior which is in contrast with
the analysis in Ref.\cite{Jabbari}.
We would like to determine the low energy coupling constant and
the $\theta$ parameter.

First, we review the S-duality of non-commutative
$U(1)$ gauge theory.\cite{Duality1,Duality2}
In Ref.\cite{SWncym} they give a field redefinition
between fields in ordinary theory and in non-commutative theory.
Let us suppose the lagrangian of the non-commutative $U(1)$ gauge theory.
We denote the field of the non-commutative theory by putting hat
like $\hat A_\mu$.
We perform a field redifinition from the gauge field
$\hat A_\mu$ to the ordinary one $A_\mu$ according to Ref.\cite{SWncym}.
In order to perform S-duality we introduce (dual) auxiliary field
$B_\mu$ for imposing the Bianchi identity $dF\equiv0$.
Then, eliminating the field strength $F$ and performing a renormalization
transformation from $B_\mu$ back to $\hat B_\mu$,
we end up with a non-commutative $U(1)$ gauge theory.
This is a dual description of the initial non-commutative theory.
In this description we find the S-duality relation\cite{Duality1}
\be
g_D = 1/g ,\quad
\theta_{Dij} = -\frac{g^2}{2}\epsilon_{ijkl}\theta_{kl} .
\ee
This argument holds in order by order of $\theta$, and
there is no full order treatment for the non-commutative S-duality.
In this paper we assume, or believe, the relevance of this S-duality
in quantum field theory.

In Ref.\cite{Duality2} it is conjectured that strongly coupled
spatially non-commutative ${\cal N}=4$ Yang-Mills theory is dual to
to a weakly coupled non-commutative open string (NCOS) theory.
One may think that this duality will hold for the ${\cal N}=2$ theory
after some modification of the theory.
In the NCOS theory the effective Regge slope parameter is given by
the non-commutative parameter as $\alpha'_{eff}=\theta/(2\pi)$.
Then, the NCOS theory reduces to a non-commutative Yang-Mills theory
when the Yang-Mills non-perturvative scale $\Lambda$ is smaller than
$1/\alpha'_{eff}$.
In this situation the above S-dualities\cite{Duality1,Duality2} will be
essentially the same.

%%%%%%%%%%%%%%%%%%%%%%%%%%%%%%%%%%%%%%%%%%%%%%%%%%%%%%%%%%%%%%%%%%%%%%%%%%%%%

\section{The non-commutative Yang-Mills Theory}\label{NCYMT}
\reseteqnum

We consider the ${\cal N}=2$ non-commutative Yang-Mills theory with
the gauge group $G=U(2)$, described by a vector multiplet.
The ${\cal N}=2$ vector multiplet contains gauge field $A_\mu$,
two Weyl fermions $\lambda_\alpha,\psi_\alpha$ and
a complex scalar $\phi$, all in the $U(2)$ adjoint representation.
In ${\cal N}=1$ superspace, the lagrangian is
\bea
g^2L &=& \frac{1}{4}\left[\int d^2\theta W^\alpha W_\alpha
     + \mbox{h.c.}\right]
     + \int d^4\theta \overline\Phi e^{-2V}\Phi e^{2V}\nn
&=& - \frac{1}{4}F^{mn}F_{mn} - D^m\overline\phi D_m\phi
    - \frac{1}{2}[\phi,\overline\phi]^2 \nn
& &  - i\overline\lambda_{\dot\alpha i}\overline\sigma^{m\dot\alpha\alpha}
      D_m\lambda_\alpha{}^i + \frac{1}{\sqrt{2}}\epsilon_{ij}\left(
  \overline\phi[\lambda^i,\lambda^j] + \mbox{h.c.}\right) ,
\eea
where all the products are non-commutative $*$-products,
we suppress the trace over the gauge group $U(2)$ and
define $\lambda^i=(\lambda,\psi)$,
$\lambda\psi \equiv \lambda^\alpha\psi_\alpha$.
The $U(2)$ generators are
$T^a = \tau^a/2$ ($a=1,2,3$) and $T^0=1/2$, where
$\tau^a$ are the Pauli matrices.

The vacuum is determined by the condition $[\phi,\overline\phi]=0$.
The vacuum expection value (VEV) $\langle\phi\rangle$ belongs to
the Cartan subalgebra $U(1)\times U(1)$ of the $U(2)$.
The diagonal part $U(1)_0$ of the $U(2)$ is not broken,
since the diagonal part $U(1)_0$ does not act on
the VEV $\langle\phi\rangle$.
The $SU(2)$ part is broken down to $U(1)_3$,
due to the VEV $\langle\phi\rangle$.
Then, the gauge symmetry becomes $U(1)_0\times U(1)_3$
in the low energy region.

The low energy theory is described by a prepotential $F$
which is a holomorphic function as explained in \cite{Seiberg}.
The low energy effective lagrangian is\cite{SWexact}
\be
L_{eff} = \frac{1}{4\pi}{\rm Im}\left[
\int d^4\theta\frac{\partial F}{\partial A^i}\overline A^i
+ \int d^2\theta \frac{1}{2}\frac{\partial^2 F}{\partial A^i\partial A^j}
  W^i_\alpha W^{\alpha j}
\right]_* ,
\ee
where $i,j = 0,3$.
The low energy symmetry $U(1)_0\times U(1)_3$ does not factor,
because of the non-commutativity of the $*$-product.
The prepotential depends on the components $A^0, A^3$
only through the form $\sqrt{(A^3)^2+(A^0)^2}$.
This is seen easily as follows.
The gauge fields of the low energy theory are
$A_\mu=T^3A^3_\mu+T^0A^0_\mu$, and the $U(1)_0\times U(1)_3$
gauge transformations are
\bea
\delta A^0_\mu &=& \partial_\mu\alpha^0
                 + \frac{1}{2}[A^3_\mu,\alpha^0]_*
                 + \frac{1}{2}[A^0_\mu,\alpha^3]_* ,\nn
\delta A^3_\mu &=& \partial_\mu\alpha^3
                 + \frac{1}{2}[A^3_\mu,\alpha^3]_*
                 + \frac{1}{2}[A^0_\mu,\alpha^0]_* .
\eea
These two gauge fields are related by this symmetry.
These two, and as well as the weak bosons, are put together
into the prepotential by virtue of the non-commutative $U(2)$ symmetry.
This way of determining a general form of $F$ is essentially described
in Ref.\cite{SWexact}.
Then, it is enough to consider the $U(1)_3$ part with putting
$A^0=0$ formally
in the effective lagrangian to determine the prepotential.
Otherwise we consider the functional form of $F$ in
$a=\sqrt{(A^3)^2+(A^0)^2}$.
In the ordinary case the non-perturbative contributions are only
due to the anti-self-dual instantons.
However, in the non-commutative case, there are also
$U(1)$ instantons\cite{NS}.
These two types of instantons produce non-perturbative corrections,
contribute to the prepotential and their roles are very symmetric.

The one-loop contribution to the coupling constant is related to
the $U(1)_R$ anomaly by the supersymmetry.
The one-loop part of the prepotential contains a logarithmic dependence
$\sim a^2\ln a^2/\Lambda^2$ to reproduce the one-loop $\beta$-function.
The $U(1)_R$ phase rotaion reveals the anomary from the prepotential
automatically.
In this case, the anomaly term takes the $F\wedge F$ form with
the $*$-product.
Then, the $U(1)_R$ is broken to ${\rm Z}_8$ and for non-zero
$\langle\phi\rangle$ the ${\rm Z}_8$ symmetry is broken to ${\rm Z}_4$
which acts trivially on the moduli space parameterized by $u$.
The discrete symmetry ${\rm Z}_2={\rm Z}_8/{\rm Z}_4$ acts
on the moduli space by $u\to-u$ as a spontaneously broken symmetry.

%%%%%%%%%%%%%%%%%%%%%%%%%%%%%%%%%%%%%%%%%%%%%%%%%%%%%%%%%%%%%%%%%%%%%%%%%%%%%

\section{Asymptotic Behavior}\label{AB}
\reseteqnum

First, let us determine the weak coupling behavior of the
non-commutative $U(2)$ theory at large $a$.
The non-commutative theory has ultraviolet divergence and
its one-loop $\beta$ function has contributions
from planar diagrams only,\cite{SWncym} that is the same situation as
the commutative theory.
The $\beta$-function is
$\beta\equiv\mu\frac{dg}{d\mu}=-\frac{g^3}{16\pi^2} 2N_c$ with $N_c=2$.

As usual, let us combine the coupling constant and theta parameter
in the form $\tau=\frac{\theta}{2\pi}+\frac{4\pi i}{g^2}$.
The low energy values of $\tau$ are related to the prepotential.
The effective coupling, which we denote as $\tau(a)$, is parameterized
by $a$ and is given by
$\tau(a)=\frac{\partial a_D}{\partial a}=\frac{\partial^2 F}{\partial^2 a}$.
We integrate the above formula obtaining
\be
a_D = \frac{2i}{\pi}(a\ln a + a) + \cdots ,\quad a = \sqrt{2u} + \cdots ,
\ee
at $u\sim\infty$.
Let us circle around the infinity $u\to u e^{2\pi i}$ ($u\sim\infty$),
then we obtain the same monodromy
\be
M_\infty = \left(\begin{array}{cc} -1 & 2 \\ 0 & -1\end{array}\right)
\ee
as that of the ordinary $SU(2)$ Yang-Mills theory.
This result is obvious since only planar diagrams contribute to the
$\beta$-function in both cases.

Next, let us show the strong coupling behavior of the theory.
Take the S-duality transformation to the $U(1)_0\times U(1)_3$
non-commutative gauge theory.
The S-dual of the low energy theory will be the $U(1)\times U(1)$
gauge theory with one hypermultiplet.
In the classical and $\theta\to0$ limit this hypermultiplet is
the 'tHooft-Polyakov monopole.
The classical monopole solution is derived in order by order of $\theta$
in Refs.\cite{HHM,GH}.
The perturbative behavior of the dual coupling constant $\tau_D$ is
determined by this dual non-commutative gauge theory.
Calculating the $\beta$-function of the non-commutative $U(1)$ theory
with one hypermultiplet, we have
\be
\tau_D = \frac{i\beta_0}{2\pi}\ln a_D + \cdots
       = - \frac{\partial a}{\partial a_D} ,
\ee
where $\beta_0=2$ is a coefficient of the one-loop $\beta$-function.
This is integrated to be
\be
a = -\frac{i\beta_0}{2\pi}(a_D\ln a_D - a_D) + \cdots .
\ee
Since the behavior of $a_D$ is not known, we assume the form
\be
a_D = c_0(u-1)^k
\ee
with some constants $c_0$ and $k$.
We can determine the value of $k$ by a ${\rm Z}_2$ symmetry
of the moduli space.
Thus, the monodromy around $u-1\to (u-1)e^{2\pi i}$ is
\be
M_{+1} = e^{2\pi ik}\left(\begin{array}{cc}
1 & 0 \\ k\beta_0 & 1
\end{array}\right) .
\ee
Since the monodromies must obey $M_{+1}M_{-1} = M_\infty$,
we obtain
\be
M_{-1} = e^{-2\pi ik}\left(\begin{array}{cc}
-1 & 2 \\ k\beta_0 & -(2k\beta_0+1)
\end{array}\right) .
\ee

Now, let us determine the value of $k$.
Electric/magnetic charge $(n_m,n_e)$ transforms under a monodromy
transformation by $(n_m,n_e)\to M^{-1}(n_m,n_e)$.
Since the charges must be integer, $M^{-1}$ should be a integer-valued
matrix.
Then, $k$ is a integer or half-integer.
Next, we expect a ${\bf Z}_2$ symmetry which interchanges
the singularities $u=\pm 1$.
This symmetry implies a similarity transformation between the two
monodromies as $M_{-1} = A M_{+1} A^{-1}$ with some matrix $A$.
This similarity condition determines $k$ to be
\be
k = - 1 .
\ee
Thus, we determine the three monodromies
\be\label{monodromies}
M_\infty = \left(\begin{array}{cc} -1 & 2 \\  0 & -1 \end{array}\right)
          ,\quad
M_{+1}   = \left(\begin{array}{cc}  1 & 0 \\ -2 &  1 \end{array}\right)
          ,\quad
M_{-1}   = \left(\begin{array}{cc} -1 & 2 \\ -2 &  3 \end{array}\right) ,
\ee
which turn out to be exactly the same as those of
the ordinary $SU(2)$ Yang-Mills theory.
We should notice that, although we obtain the same monodromies,
singular behaviors of the non-commutative theory are
completely different.
At $u\sim1$ the VEV $a_D$ diverges as $a_D \sim 1/(u-1)$
which means that monopoles get infinitely heavy mass.
Asymptotic freedom and $a_D\sim\infty$ means that the dual theory
is in a weakly coupled phase.

%%%%%%%%%%%%%%%%%%%%%%%%%%%%%%%%%%%%%%%%%%%%%%%%%%%%%%%%%%%%%%%%%%%%%%%%%%%%%

\section{Seiberg-Witten differential}
\reseteqnum

Supersymmetry implies that the section $(a,a_D)$ is a holomorphic
function of $u$.
Even in the case of the concerned non-commutative theory,
we cannot help supposing that the section may be given
in terms of a elliptic curve.
The curve should have singularities at $u=\infty,\pm1$.
Thus, one of the candidates will be a torus of the form
\be
y^2 = (x^2-1)(x-u) .\label{torus}
\ee
This is exactly the same as that in Ref.\cite{SWexact}.
Then, the significant difference from the ordinary theory should
be implemented in the Seiberg-Witten differential one-form $\lambda$.
Now, the section is given by
\be
a_D = \oint_\beta\lambda ,\quad a = \oint_\alpha\lambda .
\ee
In what follows we would like to find the condition which should be
satisfied by the one-form $\lambda$ for the non-commutative theory.

Around a singularity any curve looks like a cylinder of the form
\be
X_z: x^2+y^2=z\label{cylinder}
\ee
up to changing variables where the parameter $z$ is a function of $u$.
In the following explanation we denote a vanishing cycle by $\alpha'$
and other non-vanishing cycle by $\beta'$.
Let us consider $\alpha'$ winding on the above cylinder and
$\beta'$ intersecting with $\alpha'$.
We choose a particular intersection of the cycles
$\alpha$ and $\beta$.
When the moduli parameter $z$ circles around the origin
$z\to ze^{2\pi i}$, we have the Picard-Lefshetz formula
\be\label{PL}
\alpha'\to\alpha' ,\quad \beta'\to\beta'-\alpha' ,
\ee
which means the curve is twisted one time around the $A_1$ singularity.

Here we briefly describe the above result.
The parametrization $r\equiv x+iy$ allows us to find a identification
between $X_z$ and a cylinder $\bf C^*$.
The Milnor fibers $M_\pm$ are defined as a upper part and lower part
of the region satisfied by $|x|^2+|y|^2>\rho$ for some $\rho>|z|^2$.
The allowed regions of $r$ are
$0<|r|^2<R_+\equiv\sqrt{\rho-\sqrt{\rho^2-|z|^2}}$ for $M_-$ and
$\sqrt{\rho+\sqrt{\rho^2-|z|^2}}\equiv R_-<|r|^2<\infty$ for $M_+$,
respectively.
Then, the mappings, ${\bf C}^*\to M_\pm$, are
$r=R_+\exp(u+i\theta)$ and $r=R_-\exp(-u-i\theta)$, respectively
for $M_+$ and $M_-$, where $u>0$, $\theta\in[0,2\pi]$.
Now, let us circle around the origin $z\to ze^{2\pi i}$.
We find $r\to re^{\pi i}$, and then $M_\pm$ rotates by $\pm\pi$
in opposite direction with each other.
This result leads to the Picard-Lefshetz formula (\ref{PL}).

Let us adopt this formula to the curve (\ref{torus}).
At $u\sim\infty$ the cycle $\alpha'=\alpha$ is vanishing.
Near the singularity the curve takes the form (\ref{cylinder})
with $z=u^{-2}$ up to changing the variables as
$x\to ux$, $y\to u^{3/2}y$.
Thus, when we circle around the infinity $u\sim\infty$,
the curve is twisted $-2$ times around the singularity.
So, we have
\be
\alpha\to\alpha ,\quad \beta\to\beta-2\alpha .
\ee
In order to reproduce the monodromy $M_\infty$ the Siberg-Witten one-form
has to transform under this encircling as $\lambda\to-\lambda$.
This means that $\lambda$ has a square root singularity at $u\sim\infty$.
In the infinity $u\sim\infty$ the $\alpha$-cycle shrinks
(seen from the eyes after changing variables),
while the section $a$ diverges as $a\sim\sqrt{2u}$.
Then, $\lambda$ have to diverge as $\lambda \sim \sqrt{u}$.
This singular behavior is the same as that of the ordinary case of
the Seiberg-Witten differential $\lambda=\sqrt{\frac{u-x}{1-x^2}}dx$,
where $x\sim{\cal O}(1)$.

In the case $u\sim1$ the cycle $\alpha'=\beta$ vanishes, and
the section $a_D$ diverges as $a_D\sim1/(u-1)$ since $k=-1$.
We find $z=-(u-1)^2/2$ and then
\be
\beta\to\beta ,\quad \alpha\to\alpha-2\beta ,
\ee
where we take account of the effect of flipping the intersection
btween $\alpha'$ and $\beta'$.
Comparing this with the monodromy matrix $M_{+1}$ in
eqs.~(\ref{monodromies}), we obtain $\lambda\to\lambda$ for
$u-1\to(u-1)e^{2\pi i}$.
Then, the one-form $\lambda$ have to behave as $\lambda\sim1/(u-1)^2$
without fractional power.

In the ordinary case $\lambda$ has a singularity only at $u\sim\infty$,
whereas in the present case $\lambda$ has three singularities at
$u\sim \infty,\pm1$.
We can learn from these situations that when a perturbative picture
appears with asymptotic freedom around a point of the moduli space,
the Seiberg-Witten one-form $\lambda$ becomes singular at that point.

Up to now we do not have found a way
for completely determine the form of $\lambda$ yet.

%%%%%%%%%%%%%%%%%%%%%%%%%%%%%%%%%%%%%%%%%%%%%%%%%%%%%%%%%%%%%%%%%%%%%%%%%%%%%

\begin{flushleft}
\bf Acknowledgements
\end{flushleft}
The authour would like to thank  M.Hayakawa, S.Mizoguchi, K.Okuyama
and especially N.Ishibashi for discussions.

% References

\end{document}